# Synchrotron-based near-field photothermal microspectroscopy: Development of a quantitative nanohistology set-up with expansion of the infrared capability 2

L. Bozec[1], G. Cinque[2], M. Reading[3], H. M. Pollock[4]
[1]Faculty of Dentistry, University of Toronto, Toronto M5G 1G6, Canada
[2]Diamond Light Source Ltd, Didcot OX11 0DE, England
[3]Department of Chemical Sciences, University of Huddersfield, Huddersfield HD1 3DH, England
[4]Dept. of Physics, Lancaster University, Lancaster LA1 4YB, England

The purpose was two-fold: To explore the capability of the Diamond Synchrotron infra-red so as to include near-field photothermal microspectroscopy (PTMS); and Toward a quantitative nanohistology - investigation of scleroderma using synchrotron radiation (mu-FTIR).

With recent advances in AFM, the integration of an IR temperature-based system on an IR beamline is still a promising and unique approach. These preliminary tests of the PTMS stand-alone system with the Diamond Multimode InfraRed Imaging and Microspectroscopy beamline (MIRIAM) were successful in one beamtime.

To our knowledge, this is still among very few studies worldwide that managed to interface an AFM-resistive probe to measure an infrared spectrum using synchrotron radiation (SR) as a source. When the alignment was successful, we managed to obtain good spectra on samples that were strong infrared absorbers. However, the alignment could not be guaranteed all the time, and the signal-to-noise ratio became too low to record any meaningful spectra on samples deemed of scientific interest for this proposal. We clarified the possible improvements to PTMS instrumentation and data acquisition needed to ensure that we can collect sufficient thermal signals from very thin samples.

**Key benefits of using the photothermal approach with SR:** PTMS is a technique for characterizing individual minute regions of an inhomogeneous specimen. Even with top-down illumination, it will never compete with tip-enhanced techniques such as nano-IR in achieving super-high spatial resolution: however, PTMS has the significant advantage that you can place the probe on any surface, including thick samples, and get an image and depth information (Dai et al 2011), something no other IR technique can give. Depth profiling is possible, even, for example, where a domain is anything other than homogeneous in the perpendicular direction - in which case it is challenging to interpret photo-thermomechanical data where sub-surface structure makes an unpredictable contribution to the results obtained at the surface (Hill et al 2009).

A sample of any thickness can be studied with a thermal probe, whereas all expansion measurements need thin samples for a defined submicron spatial resolution.

Other key benefits of using a thermal probe with SR, as compared with nano-IR probe, are:
- Little or no sample preparation needed
- Very small spot size of similar flux (orders of magnitude higher flux density) to that of a conventional globar source.

For references to the current state of the art with PTMS, see Bozec et al 2002, German et al 2006, Martin et al 2007, Bentley et al 2007, Kelly et al 2008, Walsh et al 2008, Nicholson et al (2008), Grude at al 2009, Martin & Pollock 2010, Pollock 2011, H M Pollock and S G Kazarian (2014), and Kelly et at al 2011.

***Experimental details:*** We aimed to revisit an approach that had been carried out at the Daresbury synchrotron in early 2000 (Bozec 2002). For the present experiment, a stand-alone PTMS set up mounted on a Caliber AFM (Bruker Instruments) equipped with a Wollaston thermal probe was interfaced with the MIRIAM IR spectrometer. An optical interface was placed directly in the spectrometer chamber that would re-direct the focussed infrared beam at the contact point between the thermal (mounted in the AFM) and the sample. The recording of these infrared spectra was carried out using filtering and amplifying the voltage from across the Wollaston probe (\**note: the Wollaston probe acts as a thermistor operated at constant current up to 100mA*). The bandpass filter dynamic range is set up according to the speed (v) of the mirror from the Michelson interferometer. Experiments were typically run at ~2.2kHz resulting in the frequency range between 55 to 500Hz to maximize the generation of non-radiative transitions signal (heat). The digital amplifier's input and output gains were optimized to balance the level of noise recorded in the baseline of the centre-burst and the amplitude of the centre burst. Setting those gains was also impacted by the optical alignment of the beamline at the spectrometer's input and between the optical interface and the spectrometer itself. It is worth noting that harmonic contributions from the 50Hz main supply could be detected in the spectra recorded.

A standard thermally-sensitive polymeric paper was used first to detect a hot spot on the optical interface. When this spot was detected, we confirmed the rough alignment of all the optics, which permitted the detection of the IR beam at the AFM stage. The Wollaston probe was then brought into contact with that spot to produce a centre-burst signal. Finer adjustment of the optics led to a first success in recording a spectrum of the thermally-sensitive polymeric paper was obtained. Following this early success, samples of polymers, collagen, and gelatine were investigated with mitigated successes due to issues related to signal-to-noise and optical alignment. When possible, single-beam spectra were recorded by the Wollaston probe (512 co-additions at 16cm$^{-1}$ resolution). Background spectra (probe off contact) were also recorded, but the reduced signal-to-noise ratio of the recorded single-beam spectra prevented any subsequent background subtraction to obtain a clear PTMS spectrum as expected.

***Results:*** The detailed objectives of this project were as listed below, and the results are grouped accordingly under the same headings, namely:
   a) interfacing the current PTMS set-up with the IR beamline set-up, and PTMS signal level observed,
   b) signal to noise (S/N) ratio, and comparison with synchrotron radiation (SR) µ-FTIR; sample materials tested, and sequence of scans (spectra) attempted
   c) Comparison between PTMS spectra and examples from the literature
   d) Spectra from increasingly more complex biosamples
   e) Approach towards a PTMS technique for nanohistology, and defining IR markers for scleroderma

f) Correlation of MIRIAM IR data and PTMS with AFM structural and mechanical data.

g) As an addition to these objectives, as part of the data analysis, it seemed worthwhile to explore the possible merits of chemometric analysis of PTMS spectra obtained.

Outline details are summarised as follows:

a) **interfacing the current PTMS set-up with the IR beamline set-up, and PTMS signal level observed**

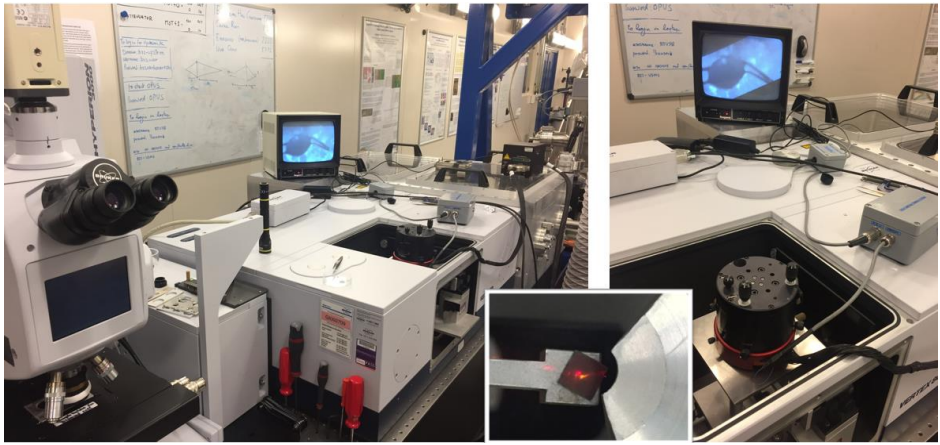

We managed to interface the stand-alone PTMS set-up successfully with the MIRIAM beamline. The optical interface that was designed for a Tensor 27(Bruker) spectrometer fitted inside the MIRIAM IR setup, i.e. a Vertex 80 V interferometer. Tthe focussed infrared beam could be diverted upwards onto the sample mounted on the interface. Using the thermally sensitive polymeric paper, we were able to observe the hot-spot indicated the focal point of the infrared beam.

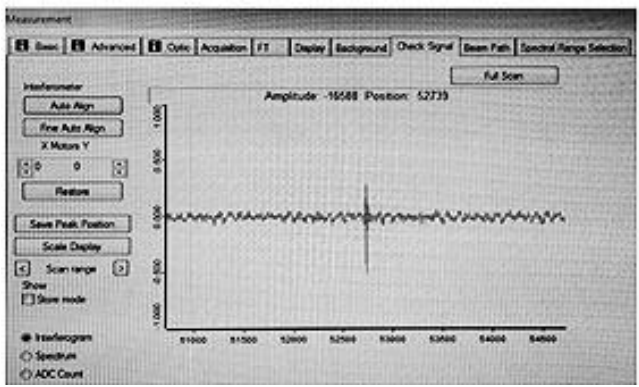
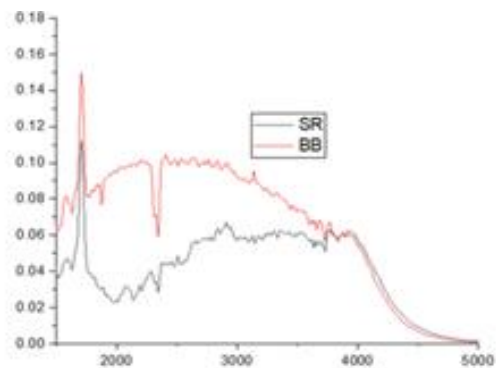

Following this, we improved the signal ratio of the centre burst recorded by the Wollaston probe brought into contact with the paper. A significant baseline noise could be recorded, including a wave pattern that suggests the presence of environmental 50Hz harmonics. These harmonics were found in the single-beam spectrum after processing. In an effort to minimize this effect, we altered the bandpass frequencies of the digital filter but without success. We also found that the amplitude of the centre-burst signal was very sensitive to the precise optical alignment of the spectrometer input and output. Much of our beamtime was spend on realigning the optics. Once the optical alignment was performed successfully, spectra could be successfully recorded, using both the broadband source and the synchrotron microbeam. At this stage, the signal level was too low for background subtraction to be worthwhile.

***Objective a) achieved***

**b) signal to noise (S/N) ratio, and comparison with synchrotron radiation (SR) μ-FTIR; sample materials tested, and sequence of scans (spectra) attempted**

Our subsequent success came from measuring the infrared spectra of the strongly absorbing infrared-sensitive polymeric paper. As shown in the figure above, very strong infrared bands can be recorded in the fingerprint region whether the source is SR (blue) or the BB (red). These spectra were recorded 16cm-1 resolution and 512 co-additions.

The signal-to-noise ratio of both these spectra is suitable for detecting strong infrared bands. It is worth noting that this sample is specially designed to absorb infrared radiation, which explains the strong signal detection. Upon repeating these measurements, we were unable to obtain such a strong thermal signal again despite efforts in realigning the beam.

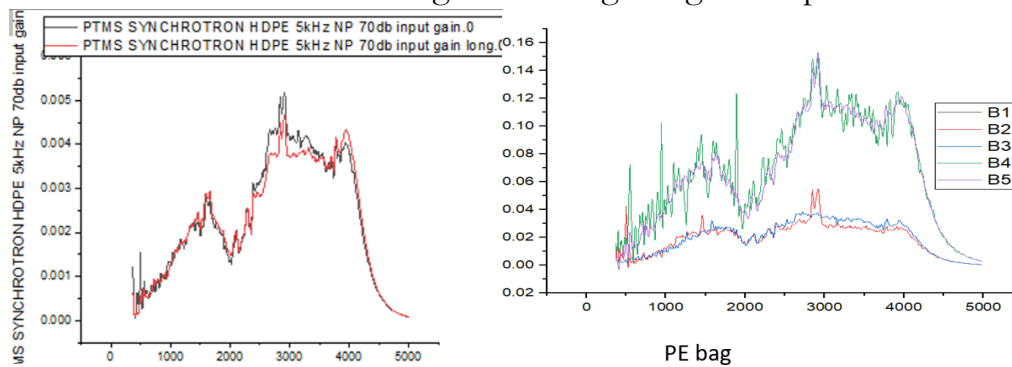

PE bag

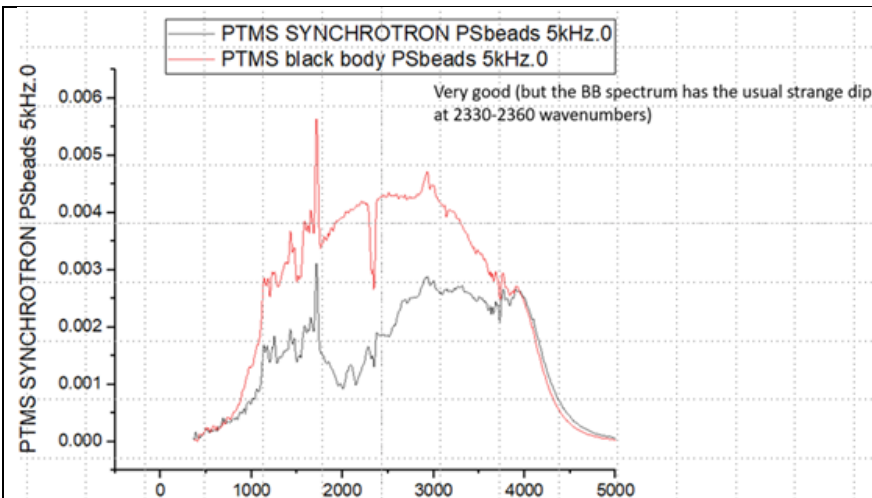

Additionally, we acquired IR signals from several sample sections of area 10x10 µm². Also, the signal output generated by the thermal probe was easily sufficient to generate an infrared spectrum. Several of the scans, as well as showing a satisfactory signal-to-noise ratio, display reasonable agreement between SR and BB spectra.

***Objective b) achieved***

**c) Comparison between PTMS spectra and examples from the literature:**

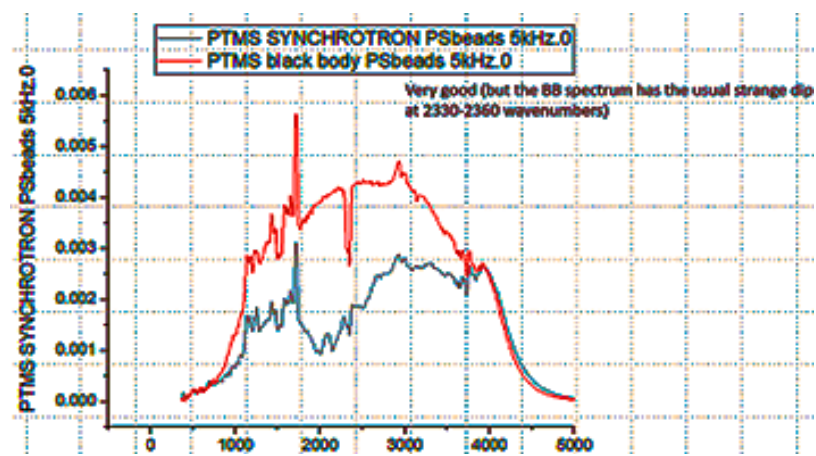

PTMS spectra from nineteen scans of COLL, gelatine, HDPE, PEbag, PSbeads, Skin, and Skinpressed were obtained. Overall, the signal-to-noise ratio in the single beam spectra recorded was insufficient to perform any background subtraction. It is, therefore, difficult to compare the recorded spectrum with literature. At this point, it is useful to examine to what extent the peaks in these spectra correspond to the values expected from spectra quoted in the published literature. In some cases, several of the spectral peaks did correspond with those of published IR spectra.

In the case of polystyrene, agreement between the baseline-corrected spectra with the literature values of peak locations is not particularly convincing. However, as shown above, we did have very good agreement between polystyrene beads spectra obtained using synchrotron radiation versus black body source. We could speculate that a possible reason for incomplete agreement between the different analysis methods used both here and in studies reported in the literature, is: an important feature of PTMS is that it samples a micron-sized area of the sample, rather than a macroscopic region.

*Objective c) partially achieved*

During the available beam time, it was decided that the following objectives could not be completed due to the relatively low SNR of the current PTMS set-up. The samples brought for these objectives mainly consisted of thin films, such as biofilms and histological sections. Considering that the thermal signal generated by thicker sample (such as the thermal paper or PS beads) yielded a low SNR and considering the challenge of the beam alignment to optimize the SNR, it was not possible to obtain useful data for the following objectives:

- Spectra from increasingly more complex biosamples
- Approach towards a PTMS technique for nanohistology, and defining IR markers for scleroderma
- Correlation of MIRIAM IR data and PTMS with AFM structural and mechanical data
- As an addition to these objectives, as part of the data analysis, it seemed worthwhile to explore the possible merits of chemometric analysis of PTMS spectra obtained.

**d) Spectra from increasingly more complex biosamples**

Here we show a preliminary result in the form of scans from skin and skinpressed samples, suggesting that the method looks promising:

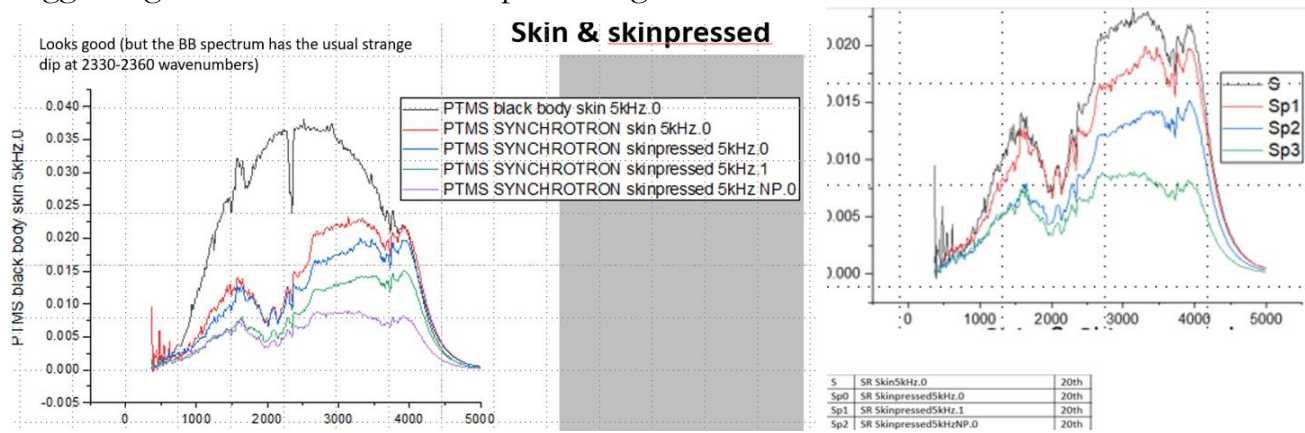

Thus we see that due to the technical challenges in terms of signal-to-noise ratio and beam alignment, only objectives a) and b) were carried out conclusively. Objective c was partially completed, and finally, objective d-e-f and g were not achieved.

*Objective (d): preliminary results only*

*Objectives (e, Approach towards a PTMS technique for nanohistology), (f, defining IR markers for scleroderma), and (g, Correlation of MIRIAM IR data and PTMS with AFM structural and mechanical data): not achieved*

**g) As part of the data analysis it seemed worthwhile to explore the possible merits of chemometric analysis of PTMS spectra obtained:**

There were not enough runs from each individual sample for principal component analysis (PCA) to be worthwhile, but nevertheless it was possible to look for clustering. Here is one of the resulting scores plots, in which PTMS spectra from samples of skin and HDPE have been compared:

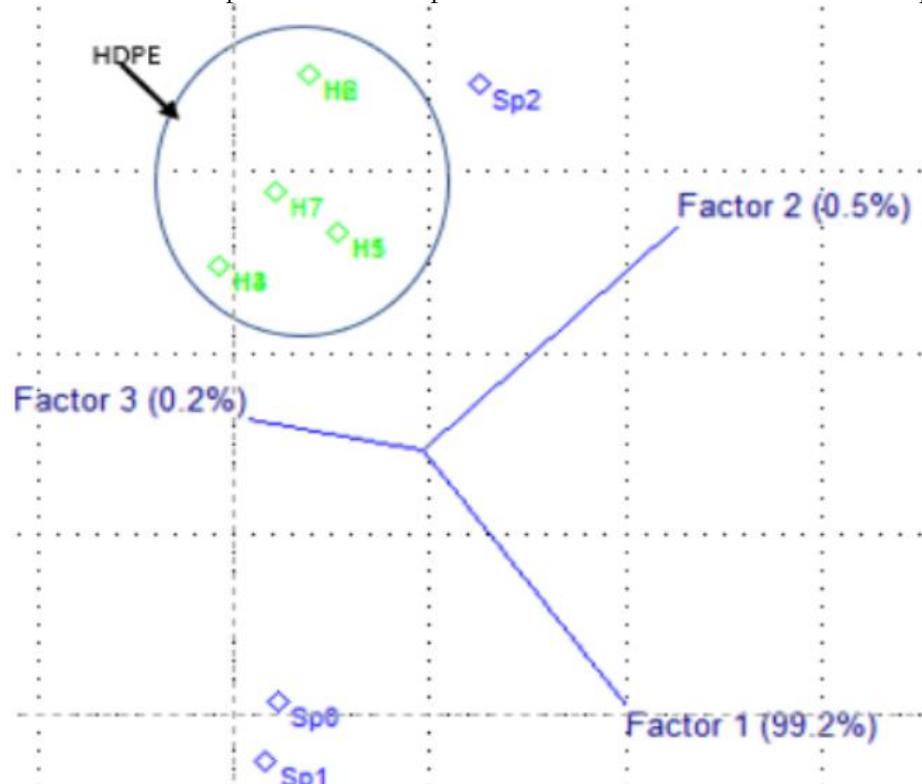

We see that the HDPE spectra prove to be the most consistent (least disparate) of all.
**Objective (g): initial work only**

**Conclusions and future work**: With recent advances in Atomic Force Microscopy, the integration of an IR AFM-based system with a synchrotron beamline is recognised as a very promising advance worldwide. Although this beamtime aimed at revisiting an approach that had been carried out at the Daresbury synchrotron in early 2000 (Bozec et al 2002), our principal new achievement consisted in proving that PTMS can be interfaced with the MIRIAM IR spectrometer. Thus the work presented here presents a significant step forward from the original Daresbury synchrotron work. Significant technical challenges remain, especially related to the optimising of the alignment, thermal signal acquisition, and finally shielding for environmental electrical noise. To our knowledge, this is still among the few studies that managed to interface an AFM/resistive probe to measure an infrared spectrum using synchrotron radiation as a source. It is worth noting here that the PTMS instrumentation was 15 years old and had remained unused for the last decade: as such, the revival of this approach is in itself a significant achievement. When the alignment was successful, we managed to obtain good spectra on samples that were good infrared absorbers. However, the alignment could not be guaranteed all the time, and the SNR became too low to record any meaningful spectra on samples deemed of interest for this proposal.

Completing the entire set of initial objectives of this proposal was an over-ambitious task, and there is now a good case for submission of a proposal to carry out the development of a quantitative nanohistology project, with expansion of the capability of the synchrotron IR station. This part of the proposal was originally aimed towards the correlation of infrared, mechanical, and structural properties of collagen on histological sections. The inability to obtain data on such samples does not mean that this cannot be done. It, however, suggests that the entire PTMS instrumentation and data acquisition approach may need to be improved to ensure that we can collect sufficient thermal signals from a very thin sample.

Further work to achieve this improvement must deal with the long data acquisition times resulting from poor signal-to-noise ratio requiring coherent averaging. This noise was attributed to three sources: (1) the detector itself: the resistive Wollaston wire generates white thermal noise, (2) external noise: electromagnetic pick-up at the level of the probe, wiring, and the amplifier – this noise shows mainly as harmonics of the mains (50 Hz), (3) amplifier noise (thermal noise, shot noise, etc.). A dedicated SPM head add-on to the IR beam station at the Diamond synchrotron radiation source should be constructed, enabling the sample to be analyzed to be positioned directly at the focal point of the infrared beam. The microscope should be designed to fit directly inside the sample chamber of most FTIR spectrometers. The rationale behind the construction of a dedicated microscope, rather than using a commercially available scanning probe microscope, lies in the importance of having the surface of the sample be scanned directly accessible from any angle, and thus to be exposed directly to the beam of the FTIR. This would avoid the need to use any optical coupling, such as IR-transmitting optical fibres, and thus eliminating energy losses and bandwidth reduction through the coupling mechanism. This open architecture approach would allow direct access for illumination of the sample, and permit a more versatile use of the instrument for many applications. One of the approaches that ought to be investigated is to include the use of one or more of the existing micromachined equivalent types of thermal probes (Hammiche et al 2000, Dekhter at al 2000, Kim et al 2007), in order to achieve the highest spatial resolution and reduce the volume of sample required to generate a thermal signal. The Wollaston probe that we used in this experiment is no longer produced.

Unfortunately, owing to the premature death of Azzedine Hammiche, it has not yet proved possible to undertake such further work.

Two more recent publications are of interest here, in connection with possible further work. Both describe work involving a tunable pulsed source rather than an SRS beamline. One study was carried out at NIST (Katzenmeyer et al. 2015). By measuring temperature (in addition to expansion), STIRM, together with AFM-IR, can provide enhanced chemical imaging capabilities for very thin samples or for samples with small thermal expansion coefficients. Preliminary data suggest that it is possible to quantitatively extract the local thermal diffusivity or even the local thermal conductivity (if $\varrho$ and C are known) from the STIRM signal, provided the temporal evolution of the signal is sample-limited. They were able to obtain a measurable signal with the SThM tip for a sample that was not measurable with AFM-IR (non-resonance-enhanced), and to extract the sample thermal conductivity.

More recently, experiments again using the IR beamline, at Diamond, together with measurements of the photothermal expansion component of the deflection signal in AFM-IR (Donaldson et al 2016), achieved a clear step forward compared with our Daresbury results (Bozec at al 2002). This RE-AFMIR approach gives a better signal-to-noise noise ratio (and hence the prospect of higher spatial resolution) than PTMS.. Details of the procedure for infrared spectral imaging at the MIRIAM IR beamline are given by Cinque et al (2017), and further developments and planned enhancements have been described subsequently (Frogley et al 2020).

We acknowledge support for time on the MIRIAM infrared beamline B22 at the Diamond Light Source Facility via academic user access route.